\documentclass[12pt,double]{iopart}
\usepackage{longtable}                  
\usepackage{dcolumn}                    
\newcolumntype{d}{D{.}{.}{-1}}          
\usepackage{natbib209}
\usepackage{lscape}
\usepackage{dcolumn} 
\usepackage{chngpage}
\usepackage{graphicx}
\newcolumntype{d}{D{.}{.}{-1}}
\include{macros}
\def\cm{~cm$^{-1}$}
\def\mic{$\mu$m}
\def\etal{et al.\/ }

\begin{document}

\title[The spectrum of Fe II]{The spectrum of Fe II}

\author{Gillian Nave$^1$, Sveneric Johansson$^2$}

\address{1. National Institute of Standards and Technology,
Gaithersburg, MD, USA.\\
2. Lund Observatory, University of Lund, Sweden.
}
\ead{gillian.nave@nist.gov}
\begin{abstract}

  The spectrum of singly-ionized iron (Fe~II) has been recorded using
  high-resolution Fourier transform and grating spectroscopy over the
  wavelength range 900~\AA\ to 5.5~$\mu$m. The spectra were observed in
  high-current continuous and pulsed hollow cathode discharges using
  Fourier transform (FT) spectrometers at the Kitt Peak National
  Observatory, Tucson, AZ and Imperial College, London and with the
  10.7~m Normal Incidence Spectrograph at the National Institute of
  Standards and Technology. Roughly 12\,900 lines were classified
  using 1027 energy levels of Fe II that were optimized to measured 
  wavenumbers. The wavenumber uncertainties of lines in the FT
  spectra range from 10$^{-4}$~cm$^{-1}$ for strong lines around
  4~$\mu$m to 0.05~cm$^{-1}$ for weaker lines around 1500~\AA. The
  wavelength uncertainty of lines in the grating spectra is 0.005~\AA.
  The ionization energy of (130\,655.4$\pm$0.4)~cm$^{-1}$ was
  estimated from the $\rm 3d^6(^5D)5g$ and $\rm 3d^6(^5D)6h$ levels.

\end{abstract}

\maketitle

\section{Preface}

The title of Sveneric Johansson's last paper, ``A half-life of Fe~II''
\citep{Johansson_09}, accurately describes his association with the
spectrum during his career. He began work on the spectrum as a graduate 
student of Edl\'{e}n, with papers on new energy levels
\citep{Johansson_74} and a comprehensive analysis of the spectrum
\citep{Johansson_78}. Much of his subsequent work focussed on the
discovery of new energy levels, identification of Fe~II lines in
astrophysical spectra, and their use for diagnostics, including the
discovery of stimulated emission from Fe~II in $\eta$ Carinae
\citep{Johansson_lasers}.  This work continued for the rest of his
life, with his final paper on new energy levels in Fe~II observable
solely in stellar spectra \citep{Johansson_09}, completed just
days before his death in 2008.

In 1987, I began working on the spectrum of Fe~I at Imperial College,
London, UK (IC), using Fourier transform (FT) spectra of iron-neon and 
iron-argon hollow cathode lamps recorded at Kitt Peak National Observatory 
(KPNO) and IC. This work resulted in two papers on precise wavelengths in 
Fe~I and Fe~II \citep{Nave_91, Nave_92}. In 1992, I went to Lund University 
and began work with Sveneric to combine the FT spectra from IC with grating
spectra recorded by him in 1988 on the 10.7-m normal incidence spectrograph 
at the National Institute of Standards and Technology (NIST). The
combined data were used to produce a linelist of about 34\,000 spectral 
lines covering wavelengths from 833~\AA\ to 5~$\mu$m. Roughly 9500 of these 
lines are due to Fe~I, and these were used to produce optimized energy 
levels and a new Fe~I multiplet table \citep{Nave_94}. About 15\,000 of the 
remaining lines belong to Fe~II. After completion of the work on Fe~I, 
I continued to collaborate with Sveneric on the spectrum of Fe~II, with the 
aim of preparing a comprehensive analysis and linelist for Fe~II.  However,
progress was slow due to competing work and his subsequent illness.
Consequently, this work remained uncompleted when Sveneric died in
October, 2008.

This paper is my best attempt to complete the analysis of Fe~II in the
way that Sveneric was unable to do. It has benefitted from 
unpublished line lists and energy levels that he gave me
while I was working in Lund and from the insights into
the spectra of iron-group elements that I gained while there.
However, the loss of Sveneric's knowledge and lack
of access to some of the original data behind the unpublished
linelists have resulted in some inconsistencies in the
analysis. In particular, it has proven to be impossible to put the
intensities on a consistent scale over the whole wavelength
region. Since I do not have the original data behind the unpublished
linelists, I was also unable to verify all of the lines in these
lists in a way that I would have done otherwise. However, these
lists appear to be reliable, and I have no reason to doubt that Sveneric prepared them
well. Although Sveneric's considerable contributions of lines and
energy levels make him a co-author of this paper, the final selection
of lines and energy levels in the tables is solely mine.

Gillian Nave

\section{Introduction}

The iron-group elements have complex spectra with thousands of lines
from the VUV to the infrared \citep{Johansson_78}. They are important 
in the absorption spectra of many astrophysical objects including the ISM, 
hot and cool stars, gaseous nebulae, galaxies and 
quasi-stellar object absorption line systems. Lines from
Fe~II in particular account for as many as half of the absorption
features in A and late B-type stars in all wavelength regions
\citep{chilupi}. They also form prominent emission lines from a wide
variety of objects including chromospheres of cool stars
\citep{Dupree_05},  active galactic nuclei \citep{Hamann_1999}, and nebular 
regions around massive stars.

Problems in modeling stellar spectra in the UV and VUV often can be traced to
the lack of adequate data for iron group elements, particularly for
very highly-excited levels of these elements.  The importance of these
highly-excited levels has been demonstrated in observations of the
Bp star HR6000 made with the HIRES spectrograph on the Very Large
Telescope.  \citet{Castelli_08} found it possible to identify strong
lines in the region around 5178~\AA\ as transitions of the
3d$^6$($^3$H) 4d-4f array due to unpublished
energy levels of Fe II.  The upper levels of this transition array are
located around 128,000 cm$^{-1}$ (15 eV) and are the highest known levels of Fe II.

Previous papers on Fe~II containing substantial numbers of lines or
energy levels are summarized in Table \ref{tab:levref}. The last
comprehensive publication on Fe~II was in 1978
\citep{Johansson_78}. It contained classification of 3272 
lines from 576 energy levels, covering the wavelength range 900~\AA\
to 11\,200~\AA. The wavelength uncertainty of the best lines was
0.02~\AA. Since then, precise wavelengths of Fe~II in the UV
\citep{Nave_91,Nave_04} and VUV \citep{Nave_97} have been
published. Additional Fe~II wavelengths have been reported by
\citet{Adam_87}, \citet{Rosberg_92}, \citet{Johansson_95},
\citet{Biemont_97}, \citet{Aldenius_06}, \citet{Aldenius_09},
\citet{Castelli_08}, and \citet{Castelli_10}. These publications
contain approximately 500 new spectral lines. Studies by
\citet{Adam_87}, \citet{Johansson_88}, \citet{Rosberg_92},
\citet{Biemont_97} and \citet{Castelli_10} account for an additional
301 new energy levels.

\begin{table}
\caption{Previous work on Fe II} \label{tab:levref}
\begin{footnotesize}
\begin{tabular}{llll}
\hline
Reference               &  No. energy levels & No. lines  & Principal configurations \\
\hline					                 
\citet{Johansson_78}    &  576               & 3272       & 3d$^6$ns, 3d$^6$np, 3d$^6$nd, 3d$^6$4f \\
		        &                    &            & 3d$^5$\,4sns,(3d$^5$4s)np, (3d$^5$4s)4d, 3d$^5$4p$^2$ \\
\citet{Adam_87}         &   20               & 120        & 3d$^5$\,4s($^7$S)4d, 3d$^5$($^6$S)4p$^2$ \\
\citet{Johansson_88}    &   73               & -          & 3d$^6$($^3$L)4d, where  L=P, F, G, H   \\
\citet{Nave_91}         &    -               & 221        &                                        \\
\citet{Rosberg_92}      &   49               & 220        & 3d$^6$($^5$D)5g                        \\
\citet{Nave_97}         &    -               & 473        &                                        \\
\citet{Biemont_97}      &   50               & 74         & 3d$^6$($^5$D)6h                        \\
\citet{Castelli_10}     & 109  (58 used)     & 137$^a$    & 3d$^6$($^3$L)4f,3d$^6$($^3$L)nd where L=P, F, G, H       \\
This work               & 102                & 12\,909    &                                        \\
\hline
\end{tabular}
$^a$ Lines listed as laboratory measurements that are not included in
\citet{Johansson_78}. 51 of the energy levels in \citet{Castelli_10}
could not be confirmed using our spectra.
\end{footnotesize}
\end{table}

\section{Experimental work}\label{experiment}

The laboratory spectra used here are the same as those 
in our previous studies of Fe~I and Fe~II
\citep{Nave_91,Nave_92,Nave_94,Nave_97}. The spectra were obtained on four 
different instruments: the f/55 IR-visible-UV FT spectrometer at the Kitt Peak
National Observatory (KPNO), Tucson, Arizona for the region 2000~\cm\ to
35\,000~\cm\ (5~$\mu$m to 2900~\AA\ ); the f/25 vacuum UV FT spectrometer at
Imperial College, London (IC) for the region
33\,000~\cm\ to 67\,000~\cm\ (3000~\AA\ to 1500~\AA); the f/25 UV FT
spectrometer at Lund University, Sweden for the region 31\,900~\cm\ to
55\,000~\cm (3135~\AA\ to 1820~\AA); and the 10.7m Normal
Incidence Grating Spectrograph at NIST for high-dispersion grating
spectra above 30\,770~\cm\ ($<$3250~\AA).

\subsection{Fourier transform spectra}\label{FTS_spectra}

The light source used for the FT investigations was a hollow cathode
lamp, run in either neon or argon. This source emits lines of Fe~I,
Fe~II, the neutral and singly-ionized spectra of the carrier gas used,
and a small number of impurities. The cathode was a 35~mm long cylinder of
pure iron with a 8~mm bore. A water cooled cathode was used
 for some of the spectrograms. The metal case of the lamp formed the
anode. The gas pressures were about 500~Pa of Ne or 400~Pa of Ar for
the visible and IR observations made at KPNO, and 300~Pa to 800~Pa of Ne for the UV
observations made at IC. The currents ranged from 320~mA to 1.1~A.  Argon-iron
spectra were recorded in the region 17\,500~\cm\ to 35\,000~\cm\ to provide an
absolute wavelength calibration based on Ar II lines. One infrared
spectrum was recorded with a water-cooled cathode as source and a
higher current of 1.4~A. The spectra are summarized in Table
\ref{tab:spectra}.

\begin{table}
\caption{Spectra used in this analysis$^a$}\label{tab:spectra}
\begin{scriptsize}
\begin{tabular}{llllllll}
\hline
Name$^b$     &  lower 
                        & upper  
                                 & Resolution &  Gas & Pressure  &  Current & Notes \\
             & wavenumber & wavenumber \\
             & (cm$^{-1}$)
                        & (cm$^{-1}$)    
                                 & (cm$^{-1}$)&      & Pa (Torr) &  A       &       \\
\hline
\multicolumn{5}{l}{{\bf KPNO FT spectrometer. CaF$_2$ beamsplitter}} \\
820216R0.001 & 2006       & 8996   & 0.012      &  Ne  & 373 (2.8) &  1.4     & Water cooled cathode; \\
	     &            &        &            &      &           &          & N6 in \citet{Nave_92} \\ 
801211R0.002 & 5007       & 5544   & 0.0095     &  Ne  & 533 (4.0) &  0.85    & N1 in \citet{Nave_92} \\  
 \\
\multicolumn      {5}{l}{{\bf KPNO FT spectrometer. Visible-UV beamsplitter}}                                 \\
800321R0.001 & 5000       & 9436   & 0.012      &	 Ne  & 533 (4.0) &  0.86    & N2 in \citet{Nave_92} \\ 
810731R0.007 & 5007       & 11\,743  & 0.015      &  Ne  & 546 (4.1) &  1.04    &                      \\ 
810731R0.009 & 10\,343    & 13\,350  & 0.020      &	 Ne  & 560 (4.2) &  1.05    & N3 in \citet{Nave_92} \\ 
810731R0.005 & 10\,505    & 13\,820  & 0.018      &	 Ne  & 546 (4.1) &  1.04    &                      \\ 
810622R0.021 & 18\,024    & 35\,984  & 0.043      &  Ne  & 533 (4.0) &  0.75    & k11 in \citet{Nave_91}; \\
	     &            &        &            &      &           &          & N2 in \citet{Learner_88}  \\
810622R0.009 & 17\,198    & 34\,040  & 0.043      &  Ar  & 400 (3.0) &  0.4     & k19 in \citet{Nave_91}; \\
	     &            &        &            &      &           &          & A1 in \citet{Learner_88}  \\
810724R0.003 & 16\,013    & 17\,950  & 0.025      &  Ar  & 413 (3.1) &  1.0     & \\ 
810724R0.002 & 20\,801    & 26\,000  & 0.036      &  Ar  & 400 (3.0) &  1.0     & \\ 
800325R0.007 & 14\,220    & 17\,060  & 0.025      &  Ar  & 400 (3.0) &  0.81    & \\ 
810622R0.012 & 17\,006    & 25\,000  & 0.033      &  Ne  & 533 (4.0) &  0.75    & N1 in \citet{Learner_88}  \\ 
 \\
\multicolumn      {5}{l}{{\bf IC FT spectrometer.}}                                                 \\
fe6          & 33\,000    & 44\,500	 & 0.07	      &  Ne  & 533 (4.0) &  0.75    & \\
i74          & 35\,000    & 46\,000  & 0.105      &  Ne  & 400 (3.0) &  0.35    & \\
i20          & 41\,500    & 48\,000	 & 0.17	      &  Ne  & 333 (2.5) &  0.95    & \\
i22          & 41\,500    & 48\,000	 & 0.17	      &  Ne  & 333 (2.5) &  0.95    & \\
i24          & 38\,000    & 44\,000	 & 0.17	      &  Ne  & 333 (2.5) &  0.95    & \\
fen180       & 44\,000    & 59\,000	 & 0.08	      &  Ne  & 800 (6.0) &  0.5     & \\
fen200       & 44\,000    & 59\,000  & 0.08       &  Ne  & 800 (6.0) &  0.5     & \\
fe7          & 56\,000    & 75\,000  & 0.07       &  Ne  &   $^c$    &  0.5     & \\
fe8          & 50\,300    & 67\,000  & 0.08       &  Ne  &   $^c$    &  0.5     & \\
\\
\multicolumn      {5}{l}{{\bf Lund FT spectrometer.}}	       					\\
fe04193      & 31\,600    & 47\,000  & 0.07       &  Ar/Ne &  $^c$    &  1.0     & \\
fe04201      & 31\,600    & 47\,000  & 0.07       &  Ne  & 67  (0.5) &  1.0     & \\
\vspace{5mm} \\
\multicolumn      {5}{l}{{\bf NIST normal incidence grating spectrograph.}}      
\end{tabular} 
\begin{tabular}{lllllll}
\hline
                   & lower  & Upper  & Gas & Pressure  & Current  & Notes \\
                   & wavelength 
                            & wavelength
                                     &     & Pa (Torr) & A \\
                   & (\AA ) & (\AA )    \\
\hline
plate 8,6$^\circ$  & 836    & 1523   & Ne  & 130 (1.0) &  0.66    & Pulsed, 100 A peak,\\
                   &        &        &     &           &          & pulse width 70 $\mu$s, 100~Hz \\
plate 8,6$^\circ$  & 836    & 1523   & Ar  &  40 (0.3) &  0.61    & Pulsed, 100 A peak,\\
                   &        &        &     &           &          & pulse width 70 $\mu$s, 100~Hz \\ 
plate 4,8$^\circ$  & 1400   & 2107   & Ne  & 130 (1.0) &  0.64    & Pulsed, 100 A peak,\\
                   &        &        &     &           &          & pulse width 60 $\mu$s, 100~Hz \\
plate 4,8$^\circ$  & 1400   & 2107   & Ar  &  40 (0.3) &  0.64    & Pulsed, 100 A peak,\\
                   &        &        &     &           &          & pulse width 60 $\mu$s, 100~Hz \\ 
\hline
\end{tabular}  
\end{scriptsize}
{\scriptsize
                                     
$^a$ Additional grating spectra were taken between 2107~\AA\ and 3249~\AA\
under similar conditions to the above spectra. Linelists for these
spectra were given to G. Nave by S. Johansson in the early 1990's but
full details are not known. Roughly 2000 Fe~II lines have been taken
from these spectra and are included in Table \ref{tab:linelist}. \\
$^b$ Name of spectrum in the National Solar Observatory digital Library \citep{NSO}.  \\ 
$^c$ The pressure was not recorded for these archival spectra.
}                                            
\end{table}                        

The wavenumber, integrated intensity, and width of all lines
in the FT spectra were obtained using the {\sc Decomp} program 
of \citet{Brault_89} and its modification {\sc Xgremlin}
\citep{Nave_97b}, which fit Voigt profiles to the spectral lines.  Each line
was measured in up to 9 different spectra. The
full width at half maximum (FWHM) of the lines varied from about
0.024~\cm\ at 5000~\cm\ (0.1~\AA\ at 2~\mic) to about 0.24~\cm\ at
50\,000~\cm\ (10~m\AA\ at 2000~\AA).

The spectra were calibrated with lines of Ar~II between
19\,429~cm$^{-1}$ and 22\,826~cm$^{-1}$ taken from
\citet{Whaling_95}. The calibration was carried into the UV and
infra-red using wide-range spectra (Nave \etal 1991, 1992). Details of
the calibration of the visible and UV spectra are given in
\citet{Nave_11}, and of the IR spectra in \citet{Nave_92}. The
calibration of the UV spectra based on Ar~II agrees to better than
1:10$^8$ with a calibration based on Mg~I and Mg~II lines measured with a frequency comb
\citep{Salumbides_06,Hannemann_06,Batteiger_09}.

The uncertainty of the wavenumber of a line measured in a single spectrum 
is a sum in quadrature of the statistical uncertainty in the measurement of its position and of the calibration uncertainty for the whole spectrum. The
statistical uncertainty was estimated from the FWHM of the line
divided by twice the signal-to-noise ratio (SNR). This was derived
from equation 9.3 of \citet{Davis_01}, assuming 4
statistically independent points in a line width. The SNR was
estimated using a global noise level for each spectrum, but was 
limited to 100 for strong lines for the purposes of calculating the
uncertainty. This accounts for an increased noise level around the
strongest lines in the KPNO spectra due to low frequency ghosts and 
also ensures that strong lines
measured in more than one spectrum receive similar weighting in the
calculation of uncertainty of the weighted mean wavenumber. The
statistical uncertainty of strong lines (SNR$>$100) varies from
0.0001~\cm\ (0.8~m\AA\ at 2~\mic) in the infra-red to 0.001~\cm\
(0.01~m\AA\ at 3000~\AA) in the ultraviolet. The weakest lines in the
spectra have a SNR of about 3, and their uncertainty varies from about
0.005~\cm\ (0.02~\AA\ at 2~\mic) in the infrared to 0.05~\cm\ (5~m\AA\
at 3000~\AA) in the ultraviolet.  For lines measured in several
spectra, a weighted average wavenumber and uncertainty were calculated
using the squared reciprocal of the statistical uncertainty as a
weight. Since the statistical uncertainties of the individual lines
are uncorrelated, their squared reciprocal was also used as a weighting
factor in the optimization of the energy levels.

The calibration uncertainty of a spectrum consists of two parts: the
uncertainty of the original standards and the uncertainty of the calibration
constant derived from those standards. The uncertainty of the original
Ar~II standards is 2x10$^{-4}$
cm$^{-1}$. These  Ar~II standards are used to calibrate a `master spectrum',
covering the wavenumber region 17\,200~cm$^{-1}$ to 34\,040~cm$^{-1}$
(810622R0.009 in Table \ref{tab:spectra}), with a one standard
uncertainty of 1.8 parts in 10$^8$. This calibration is propagated
to the UV and IR regions using overlapping spectra, each of which
increases the calibration uncertainty. The shortest UV spectrum (fe7 in Table 
\ref{tab:spectra}) thus has the largest calibration uncertainty of 4 parts
in 10$^8$. The calibration uncertainty for a particular line can thus vary, depending on
how the spectra in which it was measured were calibrated. For simplicity, we
adopted a global calibration uncertainty of 4 parts in 10$^8$ for
all spectral lines. Since this calibration uncertainty is common to
all lines, it is not included in the optimization of the energy
levels, but is added in quadrature to the uncertainties of the
optimized energy level values after the optimization.

\subsection{Grating spectra}\label{grating_spectra}

Grating spectra were recorded in the region 30\,770~\cm\ to 119\,617~\cm\
(3250~\AA\ to ~836~\AA) using iron-neon and iron-argon hollow cathode lamps.
The iron-neon hollow cathode lamp was run in pulsed mode with a peak current
of 100~A, pulse width of 70~$\mu$\/s, pulse frequency of 100~Hz, 
and a gas pressure of 130~Pa. The iron-argon hollow cathode lamp was run in
pulsed mode under similar conditions with a gas pressure of about
40~Pa. Similar spectra were recorded with a continuous hollow cathode
lamp for comparison, but the results are not presented here. Details
of the grating spectra are given in Table \ref{tab:spectra}.

Spectra of the pulsed iron-neon hollow cathode between 850~\AA\ and
2107~\AA\ were read from the photographic plates using an
automatic comparator at Lund University that produced a
signal proportional to the optical density of the plate integrated over the
full height of the slit image. This produced a file of optical density
of the recorded spectrum as a function of position along the direction
of dispersion. This file was read in to {\sc Xgremlin} and
Gaussian profiles were fitted to the spectral lines to obtain the
wavelength and peak intensity. Strong spectral lines saturate the
image on the  photographic plates and give profiles with a flat top
that cannot be fitted with a Gaussian profile. The centroid of these
lines was estimated by integrating the profile between two points
taken on either side of the line profile. These wavelengths are,
however, less reliable than those obtained by fitting the line profile
with a Gaussian.

Additional lines in the spectra listed in Table \ref{tab:spectra} are
present in linelists given to G. Nave by S. Johansson in 1992. These are all either
weak lines that are visible on the photographic plates but not easily
fit by the {\sc Xgremlin} software, or lines blended with much
stronger lines. Roughly 360 lines of this type between 850~\AA\ and 1600~\AA\ have
been included in Table \ref{tab:linelist}. Linelists for additional
spectra between 1600~\AA\ and 3250~\AA\ were given to G. Nave by S. Johansson in the
mid 1990's. These were obtained from spectra recorded under similar
conditions to the spectra in Table \ref{tab:spectra}, but full details
of those spectra are not known. Roughly 2200 lines from these spectra are included in
Table \ref{tab:linelist}.

All grating spectra were calibrated from Ritz wavelengths of Fe~II
lines derived from energy levels determined from the FT spectra, with
particular care being taken to avoid lines that were weak, were saturated
on the photographic plate, or were significantly asymmetric. The wavelength
uncertainty of these calibration lines is approximately 2~m\AA. Details of the
calibration procedure are given in \citet{Nave_97}.

\section{Line identifications}

The initial identification of the lines was performed solely in the FT
spectra. These line identifications were carefully examined to eliminate spurious
coincidences of lines and energy level differences and were then used to obtain
optimized values for many of the energy levels. The remaining 
levels were found using unidentified lines in the FT and 
grating spectra and the identifications were again examined to
eliminate spurious coincidences. Finally, both the FT and grating
spectra were used to obtain optimized energy level values and Ritz
wavelengths and wavenumbers. The full procedure is as follows.

About 28\,000 lines were measured in the FT spectra. Known lines that
belong to species other than Fe~II were identified by comparison to
previously published wavelengths in Fe~I \citep{Nave_94}, Ar~I-II
\citep{Whaling_95,Whaling_07}, Ne~I-III
\citep{Sansonetti_04,Saloman_04,Kramida_06}, 
and various impurities present in the spectra \citep{ASD}. An initial
identification of Fe~II lines was made by comparison with Ritz 
wavelengths calculated from Fe~II energy levels taken from the
references in Table \ref{tab:levref}.  Many of these levels
had large uncertainties as
they were obtained from grating spectra of lower resolution and
accuracy than our FT spectra.  Setting a large tolerance window for
the agreement between the Ritz and experimental wavelengths leads to
many spurious identifications of lines from these levels. A better
approach is to use a few lines known to combine with these levels to
obtain a better value for the energy level. The energy levels then have 
low uncertainties and the tolerance window
can be set lower, leading to fewer spurious identifications.

Roughly 10\,000 lines in the FT spectra matched energy level
differences in Fe~II.  About 1/3 of these lines had more than one
possible identification. The identifications of all lines were checked
to ensure that very few spurious identifications contributed to the
energy level optimization, as even a small number of mis-identified
lines can have a significant effect on the optimized energy
levels. This task was aided by the small uncertainty of the
wavenumbers obtained from a FT spectrometer, and by predicted
intensities from the atomic structure calculations of
\citet{Kurucz_10}. Although the accuracy of these calculations is
limited by strong configuration interaction in Fe~II, they were useful
for locating lines in FT spectra from levels that had been found using
less accurate data from grating spectrographs. After eliminating
spurious identifications, the total number of Fe~II lines in the FT
spectra was 8930, of which 798 had more than one plausible
identification.

The lines measured in FT spectra were used to derive optimized energy levels and
Ritz wavenumbers using the  computer program {\sc lopt}
\citep{Kramida_11}. Values for 942 energy levels of Fe~II were derived
from 8930 lines covering wavenumbers from 2008~cm$^{-1}$ to
67\,851~cm$^{-1}$. The line uncertainties assigned for use in the level
optimization omit the calibration uncertainty (see section
\ref{FTS_spectra}). The statistical
uncertainty, estimated as described in section \ref{FTS_spectra}, was
added in quadrature to a minimum estimated uncertainty of
0.001~cm$^{-1}$.  This value was chosen to ensure that the level
optimization was not dominated by the infrared region, where narrow
linewidths give very precise wavenumbers, but the calibration uncertainty of the
spectra is higher as many overlapping spectra are required to reach the Ar~II
standards.  Weights were then assigned proportional to the
squared reciprocal of the estimated uncertainty of the wavenumber of
the line. Lines with more than one possible classification, lines that
were blended, and lines with a large difference between the observed
and Ritz wavenumber were assigned a low weight so that they did 
not significantly affect the values of the optimized energy levels.

The optimization was performed in three different steps. The first step
was designed to obtain accurate values and uncertainties for the ground 
term, a$^6$D. The values for the a$^6$D intervals can be 
determined from differences between lines close to one another in the
same spectrum sharing the same calibration, hence the calibration
uncertainty does not contribute to the uncertainty in the relative
values of these energy levels. An optimization was thus performed with
a set of lines connecting the lowest a$^6$D term to higher 
$\rm 3d^6\,(^5D)4p$ levels. These lines were assigned a
weight proportional to the squared reciprocal of the statistical
uncertainty of the wavenumber.  The minimum estimated
uncertainty of 0.001~cm$^{-1}$ was omitted as all the lines are in the 
same spectral region. In the second step, the a$^6$D
levels were fixed to the values and uncertainties determined from the
first step. The weights of all the lines in the FT spectra were assigned by
combining in quadrature the statistical uncertainty and the minimum
estimated uncertainty of 0.001~cm$^{-1}$ in order to obtain accurate
uncertainties for the $\rm 3d^6\,(^5D)4p$ and higher levels.

After the second step of the level optimization, a further 86
energy levels that had not been optimized with the FT spectra were
added. Most of these were from \citet{Biemont_97} and
\citet{Castelli_10}. Roughly half of the levels in \citet{Castelli_10}
could not be matched definitively to lines in the FT spectra. Levels
were adopted from \citet{Castelli_10} if they had strong transitions
that matched at least two lines in the FT spectra. All the energy
levels were then used to identify additional lines in both the FT and
grating spectra. The final optimization step derived values for 1027 energy
levels from 13\,653 transitions in both FT and grating spectra, again fixing 
the values of the a$^6$D levels to the values obtained in the first step.   
An uncertainty of 0.005~\AA\ was assigned to all grating lines. The energy 
level uncertainties from this iteration were added in quadrature to a global 
calibration uncertainty of 4x10$^{-8}$ times the value of the energy level.

\section{Energy Levels and Lines}

The full table of 1027 energy levels is available online and a small
section is given in Table \ref{tab:enlevs}. 
The configurations and terms in columns 1 and 2 are taken from the
NIST Atomic Spectra Database \citep{ASD}, the papers in Table
\ref{tab:levref}, or the calculations of 
\citet{Kurucz_10}. Many of the levels between 114$\,$212 cm$^{-1}$ and
114$\,$673~cm$^{-1}$ have not been assigned to configurations. The levels
in this region are from the two overlapping configurations
3d$^6$($^5$D)6d and 3d$^6$($^3$D)4d. The energy level values are given
in column 4. The uncertainties with respect to the ground term are
given in column 5 and represent one standard uncertainty. They were obtained
from the {\sc lopt} program by combining the uncertainties derived
from the level optimization in quadrature with a global calibration
uncertainty of 4$\times$10$^{-8}$ times the level value. The last
column lists the number of observed lines that combine with the level.

\begin{table}
\caption{Energy levels of Fe II (full table available online) }\label{tab:enlevs}
\begin{tabular}{rrrddr}
\hline
Assigned
      &Term
      &J
      & \multicolumn{1}{c}{Level}
      & \multicolumn{1}{c}{Unc.$^a$ }
      & No.              \\
       Configuration
      &
      &
      &  \multicolumn{1}{c}{($\mathrm{cm^{-1}}$) }
      &  \multicolumn{1}{c}{($\mathrm{cm^{-1}}$) }
      & lines                       \\
 \hline
 $\mathrm{3d^6(^5D)4s }$  &    $\mathrm{a^6D}$ &  9/2 & 0.0000     & 0.0000  & 42 \\ 
                          &                    &  7/2 & 384.7872   & 0.0003  & 60 \\ 
                          &                    &  5/2 & 667.6829   & 0.0003  & 64 \\ 
                          &                    &  3/2 & 862.6118   & 0.0004  & 52 \\ 
                          &                    &  1/2 & 977.0498   & 0.0004  & 30 \\ 
        $\mathrm{3d^7 }$  &  $\mathrm{ a^4F }$ &  9/2 & 1872.5998  & 0.0006  & 63 \\ 
                          &                    &  7/2 & 2430.1369  & 0.0006  & 77 \\ 
                          &                    &  5/2 & 2837.9807  & 0.0007  & 76 \\ 
                          &                    &  3/2 & 3117.4877  & 0.0008  & 54 \\ 
 $\mathrm{3d^6(^5D)4s }$  &    $\mathrm{a^4D}$ &  7/2 & 7955.3186  & 0.0007  & 65 \\ 
                          &                    &  5/2 & 8391.9554  & 0.0007  & 72 \\ 
                          &                    &  3/2 & 8680.4706  & 0.0007  & 58 \\ 
                          &                    &  1/2 & 8846.7837  & 0.0008  & 32 \\ 
        $\mathrm{3d^7 }$  &  $\mathrm{ a^4P }$ &  5/2 & 13474.4474 & 0.0009  & 67 \\ 
                          &                    &  3/2 & 13673.2045 & 0.0010  & 63 \\ 
                          &                    &  1/2 & 13904.8604 & 0.0012  & 38 \\ 
                          &  $\mathrm{ a^2G }$ &  9/2 & 15844.6485 & 0.0012  & 55 \\ 
                          &                    &  7/2 & 16369.4098 & 0.0013  & 59 \\ 
                          &  $\mathrm{ a^2P }$ &  3/2 & 18360.6399 & 0.0016  & 42 \\ 
                          &                    &  1/2 & 18886.773  &  0.002  & 29 \\ 
                          &  $\mathrm{ a^2H }$ & 11/2 & 20340.2461 & 0.0013  & 43 \\ 
                          &                    &  9/2 & 20805.7632 & 0.0014  & 57 \\ 
                          & $\mathrm{ a^2D2 }$ &  5/2 & 20516.9534 & 0.0016  & 62 \\ 
                          &                    &  3/2 & 21307.999  &  0.002  & 47 \\ 
$\mathrm{3d^6(^3P2)4s }$  &  $\mathrm{b^4P  }$ &  5/2 & 20830.5534 & 0.0011  & 75 \\ 
                          &                    &  3/2 & 21812.0454 & 0.0012  & 61 \\ 
                          &                    &  1/2 & 22409.8178 & 0.0013  & 47 \\ 
  \\  \hline
$^a$ One standard uncertainty \\
\end{tabular}
\end{table}

A small section of the table of observed Fe~II lines is given in Table
\ref{tab:linelist}. The full table of 13\,653 transitions is available
online. The intensities in column 1 depend on the source conditions
and method of measurement. The wavenumbers and intensities of all of
the lines above 3250~\AA\ were taken from a weighted average of up to
nine individual measurements in the FT spectra.  Since these
intensities were measured using several different source conditions,
they are useful only as a guide to the approximate strength of the
line and should not be relied upon for accurate intensity
measurements. Wavenumbers and intensities of lines below 3250~\AA\
that are marked with an `F' in column 11 were also taken from FT
spectra. The wavelengths and intensities of lines marked `G' in column
11 are from the grating spectra of the pulsed iron-neon hollow cathode
lamp listed in Table \ref{tab:spectra}. The intensity scale of the
grating lines is different from the intensity scale of the FT spectra,
and are given as decimal numbers in Table \ref{tab:linelist}. Strong
lines that saturate the photographic plate are indicated with a 'S' in
column 11. Wavelengths and intensities of lines taken from the
additional linelists are indicated with `N' in column 11 if taken from
the pulsed iron-neon hollow cathode, `A' if taken from the pulsed
iron-argon hollow cathode lamp. The intensities are visual estimates
of the photographic density and are also given as decimal numbers,
ranging from 0 to 5 for most lines.  The intensity scale is different
from both the lines in the FT spectra and the grating spectra lines
marked with a `G'. Lines marked `b' are broad, those marked `0d' are
faint and diffuse and those marked `0d?'  are hardly detectable from
the background.  Lines below 1600~\AA\ taken from the linelists given
to G. Nave by S. Johansson have uncertain intensities as they are all
weak or blended. They have been assigned an intensity of ``00'' and
indicated with `L' in column 11.

The wavelength in column 2 was derived from the wavenumber in column 4 for
lines taken from the FT spectra. Air wavelengths, given for lines
between 2000~\AA\ and 2~$\mu$m, were derived from the 5 parameter formula in
equation 3 of \citet{Peck_72}. The wavelengths were measured directly in the
grating spectra and the wavenumbers derived from them. The corresponding one
standard uncertainties in columns 3 and 5 include contributions from
the statistical uncertainty in the measurement of the position of the
line and the calibration uncertainty for the spectrum. The one standard 
uncertainty of lines taken from grating spectra has been estimated at 
0.005~\AA. Ritz wavelengths and their statistical uncertainties were obtained 
from {\sc lopt} and are given in columns 6 and 7 respectively. The 
uncertainties were derived by adding the uncertainties from {\sc lopt} to a
global calibration uncertainty of 4$\times$10$^{-8}$. The classification of
the line is given in column 10, with the values of the lower and upper energy 
levels in columns 8 and 9 respectively. In addition to the codes indicating the
source of the data, column 11 also indicates if there are any other
identifications for the line.

\section{Ionization Energy}

\citet{Johansson_78} obtained an estimate of 130\,563$\pm$10~cm$^{-1}$
for the ionization energy of Fe~II by using a two parameter fit to the highest
J-value levels of the lowest three $\rm 3d^6(^5D)ns$ configurations
and comparing the value to similar terms in other singly-ionized
iron-group elements. A better estimate can be obtained by
using highly-excited levels that are well-described by the J$_c$K
coupling scheme. 

\begin{figure}
\begin{center}
	\includegraphics[width=100mm,angle=270]{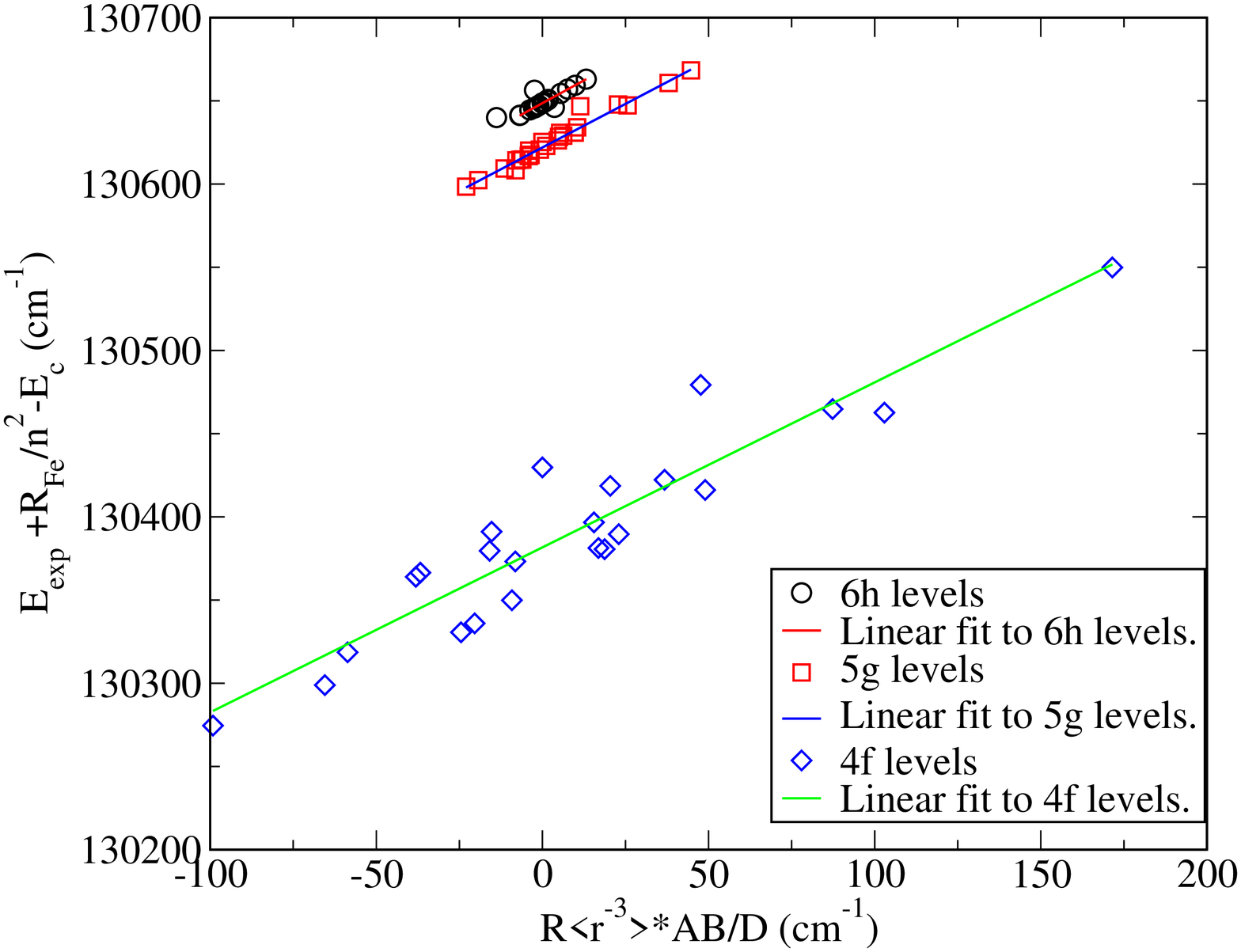}
\caption{Calculation of the ionization energy for Fe~II from the 
$\rm 3d^6(^5D)4f$,  $\rm 3d^6(^5D)5g$, and $\rm 3d^6(^5D)6h$
levels. The intercept for each configuration is $\mathrm IE-\alpha
R_{Fe}Z_c^2<r^{-4}>_{nl}$. 
\label{IP_fig}}
\end{center}
\end{figure}

We have used the quadrupole-polarization model of
\citet{Schoenfeld_95} to obtain the ionization energy from the $\rm
3d^6(^5D)4f$,  $\rm 3d^6(^5D)5g$, and $\rm 3d^6(^5D)6h$ levels. The outer
electron in these levels is weakly bound to the core and the
levels form five groups, separated by the fine structure intervals of the $\rm
3d^6\,^5D$ term of Fe III. Within each group the levels form pairs and the energy of the
center of gravity of the pairs,  $E(nlJ_cK)$, can be described by:
\begin{equation}
E(nlJ_cK) = IE +E(J_c)- R_{Fe}Z_c^2(1/n^2+\alpha <r^{-4}>_{nl}-\frac{AB}{D} Q<r^{-3}>_{nl})
\end{equation}
where  $IE$ is the ionization energy, $E(J_c)$ is the energy of the $\rm
3d^6\,^5D_{J_c}$ level in Fe III, $R_{Fe}$ is the Rydberg constant for 
Fe II (109\,736.248 cm$^{-1}$), $Z_c$ is the effective charge of the core 
(2 for Fe~II), $<r^{-3}>_{nl}$ and $<r^{-4}>_{nl}$ are hydrogenic
radial expectation values in atomic units given by equations 3 and 4 of
\citet{Schoenfeld_95}, and $A$, $B$, and $D$ are given by equation 5 of
\citet{Schoenfeld_95}. The dipole polarizability $\alpha$ of the core
and its quadrupole moment Q in atomic units are obtained from the
experimental energy levels. By plotting 
$E(nlJ_cK)+ R_{Fe}Z_c^2/n^2-E(J_c)$ for each configuration against 
$\mathrm R_{Fe}Z_c^2\frac{AB}{D} <r^{-3}>_{nl}$, a straight line of slope Q and 
intercept $\mathrm IE-\alpha R_{Fe}Z_c^2<r^{-4}>_{nl}$ is obtained.
This is shown in Fig. \ref{IP_fig}. The
ionization energy and $\alpha$ can then be obtained by simultaneously solving 
the equations for the intercepts for pairs of configurations.

The values of Q obtained from the slopes in Figure \ref{IP_fig} are 
0.99$\pm$0.08~ea$_o^2$ (where e is the elementary charge and a$_o$ 
the Bohr radius) for the $\rm 3d^6(^5D)4f$ levels, 1.05$\pm$0.02~ea$_o^2$ for the
$\rm 3d^6(^5D)5g$ levels, and 1.098$\pm$0.008~ea$_o^2$ for the $\rm 3d^6(^5D)6h$
levels. By solving the two equations for the intercepts for the 4f and 
5g levels, an ionization energy of 130\,660$\pm 5$~cm$^{-1}$ is obtained with  
$\alpha$ = 17.0$\pm$0.3. For the 5g and 6h levels, the ionization energy
obtained is  130\,655.4$\pm$0.4~cm$^{-1}$ with $\alpha$ = 15.0$\pm$0.2~a$_o^3$.
The 4f levels are
not well described by the quadrupole-polarization model as the centers
of gravity of the pairs show a large scatter around a straight line as
shown in Fig. \ref{IP_fig}. We thus recommend the value
130\,655.4$\pm$0.4~cm$^{-1}$ obtained from the 5g and 6h levels.

\section{Summary}

Table \ref{tab:linelist} contains 12\,909 spectral lines from 13\,653
transitions in Fe II, over a factor of three more than the number
of previously published lines in Fe~II. About 900 of these lines have
alternate plausible identifications due to Fe~I, Fe~III, Ne, 
Ar, or other Fe~II transitions. These lines come from a total list of about 37\,000
lines. Roughly 11\,500 of the lines in this total list remain
unidentified, and about 460 of these lines are present in the FT spectra
with a SNR $>$ 50. About half of these strong unidentified lines are
in the infrared region and are probably due to
highly-excited levels in Fe~I or Fe~II that have yet to be identified.
The uncertainties of the strongest lines in Table \ref{tab:linelist}
vary from 0.0001~\cm\ in the infrared to 0.001~\cm\ in the
ultraviolet, and are more than an order of magnitude lower than the
best lines in \cite{Johansson_78}.

\clearpage
\begin{landscape}
\begin{table}
\caption{Observed lines of Fe II (full table available online) }\label{tab:linelist}
\centering
\begin{adjustwidth}{-0.6in}{-0.5in}
\begin{scriptsize}
\begin{tabular}{rddddddddr@{ }r@{ - }r@{ }rl}
\\ \hline  Intensity$^a$ 
      & \multicolumn{1}{c}{ Vacuum }          
      & \multicolumn{1}{c}{Unc.$^b$   }       
      & \multicolumn{1}{c}{Wavenumber}    
      & \multicolumn{1}{c}{Unc.$^b$       }   
      & \multicolumn{1}{c}{Ritz Vacuum }      
      & \multicolumn{1}{c}{Unc.$^b$         } 
      & \multicolumn{1}{c}{Lower level      } 
      & \multicolumn{1}{c}{Upper level      } 
      & \multicolumn{4}{c}{Classification } 
      & Notes$^c$ \\           
      &   \multicolumn{1}{c}{wavelength}  & & & & \multicolumn{1}{c}{wavelength} \\ 
      & \multicolumn{1}{c}{(\AA\ )   }                      
      & \multicolumn{1}{c}{(\AA\ )    }                     
      & \multicolumn{1}{c}{($ \mathrm{cm^{-1}}$) }           
      & \multicolumn{1}{c}{($  \mathrm{cm^{-1}}$) } 
      & \multicolumn{1}{c}{($ \mathrm{\AA\ }$) }  
      & \multicolumn{1}{c}{($ \mathrm{\AA\ }$) }  
      & \multicolumn{1}{c}{($  \mathrm{cm^{-1}}$) } 
      & \multicolumn{1}{c}{($  \mathrm{cm^{-1}}$) } 
      & \multicolumn{3}{c}{                     }     \\  
\\ \hline                              
   0.9 &  1824.980     & 0.005    & 54\,795.13     & 0.06     & 1824.979\,49   & 0.000\,11  &  20\,805.7632 &  75\,600.900  &  $\mathrm{3d^7 }$  & $\mathrm{ a^2H }$$_{9/2 }$ &  $\mathrm{3d^6(^3D)4p }$  & $\mathrm{  w^2F^{\circ}}$$_{7/2 }$ & G    \\ 
   0   &  1824.860     & 0.005    & 54\,798.73     & 0.15     & 1824.860\,52   & 0.000\,08  &  67\,516.328  & 122\,315.037  &  $\mathrm{3d^6(^3G)4p }$  & $\mathrm{ y^2H^{\circ} }$$_{11/2 }$ &  $\mathrm{3d^6(^3H)6s }$  & $\mathrm{^2H }$   $_{11/2 }$ & N    \\ 
   0   &  1823.929     & 0.005    & 54\,826.70     & 0.15     & 1823.929\,60   & 0.000\,12  &  18\,360.6399 &  73\,187.318  &  $\mathrm{3d^7 }$  & $\mathrm{ a^2P }$$_{3/2 }$ &  $\mathrm{3d^6(^3D)4p }$  & $\mathrm{ y^2P^{\circ}}$$_{1/2 }$ & N    \\ 
    4  &  1823.8711    & 0.0008   & 54\,828.44     & 0.02     & 1823.869\,85   & 0.000\,10  &  18\,360.6399 &  73\,189.114  &  $\mathrm{3d^7 }$  & $\mathrm{ a^2P }$$_{3/2 }$ &  $\mathrm{3d^6(^3D)4p }$  & $\mathrm{ y^2P^{\circ}}$$_{3/2 }$ & F    \\ 
   1.0 &  1822.196     & 0.005    & 54\,878.83     & 0.06     & 1822.189\,74   & 0.000\,08  &   8\,680.4706 &  63\,559.498  &  $\mathrm{3d^6(^5D)4s }$  & $\mathrm{a^4D}$$_{3/2 }$ &  $\mathrm{3d^6(^3F2)4p }$  & $\mathrm{x^4D^{\circ} }$$_{1/2 }$ & G  II\\ 
   1.0 &  1822.196     & 0.005    & 54\,878.83     & 0.06     & 1822.190\,10   & 0.000\,19  &  44\,929.532  &  99\,808.549  &  $\mathrm{3d^6(^1F)4s }$  & $\mathrm{  c^2F}$$_{5/2 }$ &  $\mathrm{3d^6(^1G1)4p }$  & $\mathrm{^2G^{\circ} }$$_{7/2 }$ & G  II\\ 
   1.3 &  1822.135     & 0.005    & 54\,880.67     & 0.15     & 1822.123\,38   & 0.000\,08  &   8\,391.9554 &  63\,272.981  &  $\mathrm{3d^6(^5D)4s }$  & $\mathrm{a^4D}$$_{5/2 }$ &  $\mathrm{3d^6(^3F2)4p }$  & $\mathrm{x^4D^{\circ} }$$_{5/2 }$ & S    \\ 
    4  &  1820.9161    & 0.0008   & 54\,917.41     & 0.03     & 1820.916\,18   & 0.000\,08  &  54\,232.201  & 109\,149.611  &  $\mathrm{3d^5\,4s^2 }$  & $\mathrm{b^4G }$$_{11/2 }$ &  $\mathrm{3d^5(^2I)\,4s4p(^3P) }$  & $\mathrm{^2I^{\circ}}$$_{13/2 }$ & F    \\ 
   0.9 &  1820.480     & 0.005    & 54\,930.58     & 0.06     & 1820.478\,44   & 0.000\,17  &  31\,999.049  &  86\,929.664  &  $\mathrm{3d^7 }$  & $\mathrm{ b^2F }$$_{7/2 }$ &  $\mathrm{3d^6(^3P1)4p }$  & $\mathrm{v^4D^{\circ} }$$_{7/2 }$ & G    \\ 
   1.0 &  1819.643     & 0.005    & 54\,955.82     & 0.06     & 1819.644\,13   & 0.000\,14  &  31\,811.814  &  86\,767.614  &  $\mathrm{3d^7 }$  & $\mathrm{ b^2F }$$_{5/2 }$ &  $\mathrm{3d^6(^3P1)4p }$  & $\mathrm{v^4D^{\circ} }$$_{5/2 }$ & G    \\ 
    9  &  1818.5202    & 0.0003   & 54\,989.767    & 0.010    & 1818.521\,51   & 0.000\,08  &   7\,955.3186 &  62\,945.045  &  $\mathrm{3d^6(^5D)4s }$  & $\mathrm{a^4D}$$_{7/2 }$ &  $\mathrm{3d^6(^3F2)4p }$  & $\mathrm{x^4D^{\circ} }$$_{7/2 }$ & F *    \\ 
   0   &  1816.086     & 0.005    & 55\,063.48     & 0.15     & 1816.085\,38   & 0.000\,08  &  46\,967.4751 & 102\,030.965  &  $\mathrm{3d^6(^5D)4p }$  & $\mathrm{z^4P^{\circ} }$$_{5/2 }$ &  $\mathrm{3d^6(^5D)6s }$  & $\mathrm{^6D }$$_{7/2 }$ & N    \\ 
   1.3 &  1815.766     & 0.005    & 55\,073.17     & 0.15     & 1815.765\,92   & 0.000\,08  &   8\,391.9554 &  63\,465.134  &  $\mathrm{3d^6(^5D)4s }$  & $\mathrm{a^4D}$$_{5/2 }$ &  $\mathrm{3d^6(^3F2)4p }$  & $\mathrm{x^4D^{\circ} }$$_{3/2 }$ & S    \\ 
    4  &  1815.4116    & 0.0005   & 55\,083.927    & 0.016    & 1815.410\,97   & 0.000\,11  &  20\,516.9534 &  75\,600.900  &  $\mathrm{3d^7 }$  & $\mathrm{ a^2D2 }$$_{5/2 }$ &  $\mathrm{3d^6(^3D)4p }$  & $\mathrm{  w^2F^{\circ}}$$_{7/2 }$ & F    \\ 
  0d?  &  1810.261     & 0.005    & 55\,240.66     & 0.15     & 1810.262\,75   & 0.000\,11  &  54\,904.241  & 110\,144.841  &  $\mathrm{3d^6(^3F1)4s }$  & $\mathrm{d^2F  }$$_{7/2 }$ &  $\mathrm{3d^6(^3F)5p }$  & $\mathrm{ ^4F^{\circ}}$$_{7/2 }$ & N    \\ 
   0   &  1810.117     & 0.005    & 55\,245.06     & 0.15     & 1810.110\,58   & 0.000\,11  &  61\,093.406  & 116\,338.649  &  $\mathrm{3d^6(^3P2)4p }$  & $\mathrm{z^2D^{\circ} }$$_{5/2 }$ &  $\mathrm{3d^5\,4s(^7S)4d }$  & $\mathrm{ ^6D }$$_{5/2 }$ & N  Ne\\ 
  0d?  &  1809.945     & 0.005    & 55\,250.31     & 0.15     & 1809.944\,8    & 0.001\,0   &  50\,157.475  & 105\,407.78   &  $\mathrm{3d^6(^3F1)4s }$  & $\mathrm{c^4F  }$$_{9/2 }$ &  $\mathrm{3d^5(^2D)\,4s4p(^3P) }$  & $\mathrm{^4F^{\circ}}$$_{9/2 }$ & N    \\ 
    6  &  1809.3168    & 0.0008   & 55\,269.48     & 0.02     & 1809.318\,25   & 0.000\,11  &  21\,307.999  &  76\,577.436  &  $\mathrm{3d^7 }$  & $\mathrm{ a^2D2 }$$_{3/2 }$ &  $\mathrm{3d^6(^1S2)4p }$  & $\mathrm{x^2P^{\circ} }$$_{1/2 }$ & F    \\ 
    2. &  1809.291     & 0.005    & 55\,270.28     & 0.15     & 1809.288\,12   & 0.000\,08  &  42\,237.0575 &  97\,507.414  &  $\mathrm{3d^6(^5D)4p }$  & $\mathrm{z^6F^{\circ}}$$_{7/2 }$ &  $\mathrm{3d^6(^3P2)5s }$  & $\mathrm{ ^4P  }$$_{5/2 }$ & N    \\ 
\\ \hline

\multicolumn{14}{l}{$^a$ Intensities of the lines with different codes in `Notes' column
are on different scale and should not be relied on for accurate intensity measurements. }\\
\multicolumn{14}{l}{Decimal numbers are from grating spectra. Integer numbers are from FT spectra, which are in general the 
stronger lines.} \\
\multicolumn{14}{l}{Intensities labeled `00' below 1600~\AA\ are from unpublished linelists.
Intensities labeled `d' are diffuse; those marked `d0?' are faint and diffuse; those marked `b' 
are broad.}
\\
\multicolumn{14}{l}{$^b$ One standard uncertainty of preceding column.} \\
\multicolumn{5}{l}{$^c$ *: Line given low weight in level optimization;} \\
\multicolumn{5}{l}{\hspace*{1ex}     F: Line measured in FT spectra;} \\                      
\multicolumn{5}{l}{\hspace*{1ex}     G: Line measured in grating spectra;} \\ 
\multicolumn{5}{l}{\hspace*{1ex}     N: Line is from unpublished iron-neon linelists (see section 3.2).} \\
\multicolumn{5}{l}{\hspace*{1ex}     A: Line is from unpublished iron-argon linelists (see section 3.2).} \\
\multicolumn{5}{l}{\hspace*{1ex}     L: Line is from unpublished linelists below 1600~AA (see section 3.2).} \\ 
\multicolumn{5}{l}{\hspace*{1ex}     S: Line measured in grating spectra and is saturated on plate.} \\
\multicolumn{5}{l}{\hspace*{1ex}     I, II, III : Line blended with Fe~I, Fe~II, or Fe~III. }\\
\multicolumn{5}{l}{\hspace*{1ex}     Ne, Ar, Si, Ti, Cr, Co, Ni  : Line blended with species indicated.} \\ 
\multicolumn{5}{l}{\hspace*{1ex}     gh  : Line possibly blended with ghost. } \\      
\end{tabular}
\end{scriptsize}
\end{adjustwidth}
\end{table}
\end{landscape}

\section*{Acknowlegments}

G. Nave thanks Craig J. Sansonetti for many invaluable discussions on
wavelength calibration, spectral analysis and on the best way to
approach and present a project of this size. She also thanks Robert L. Kurucz and
Alexander E. Kramida for their assistance in finding errors in the tables. James W. Brault, Richard
C. M. Learner, Victor Kaufman, and Anne P. Thorne took or assisted
with taking many of the spectra used in this paper. Some of these
spectra are available in the National Solar Observatory digital
Library \citep{NSO}. This work was partially supported by the National
Aeronautics and Space Administration under the inter-agency agreement
NNH10AH38I.

\section*{References}

\clearpage
\pagestyle{plain}
\footnotesize                           
\setcounter{table}{2}                   
\begin{longtable}{rrrddr}   
\caption[Energy levels of Fe~II (continued)]{Energy levels of Fe II} \\ 
\\ \hline  Assigned                   
       &Term                             
       &J                                
      & \multicolumn{1}{c}{Level}       
      & \multicolumn{1}{c}{Unc. }       
      & No.              \\            
       Configuration                     
      &                                  
      &                                  
      &  \multicolumn{1}{c}{($\mathrm{cm^{-1}}$) }          
      &  \multicolumn{1}{c}{($\mathrm{cm^{-1}}$) }          
      & lines                       \\  
\\ \hline                             
 \endfirsthead 
 \multicolumn{6}{l} { \tablename\ \thetable{} Energy levels of Fe~II  -- continued from previous page} \\ 
\\ \hline  Assigned                   
       &Term                             
       &J                                
      & \multicolumn{1}{c}{Level  }     
      & \multicolumn{1}{c}{Unc.   }     
      & No.              \\            
       Configuration                     
      &                                  
      &                                  
      &  \multicolumn{1}{c}{($\mathrm{cm^{-1}}$) }          
      &  \multicolumn{1}{c}{($\mathrm{cm^{-1}}$) }          
      & lines                       \\  
\\ \hline                             
 \endhead 
\hline \multicolumn{6}{r}{{Continued on next page}} \\ 
\endfoot 
\hline 
\endlastfoot 
                          $\mathrm{3d^6(^5D)4s }$  &           $\mathrm{a^6D}$ &  9/2 & 0.0000        & 0.0000  &   42 \\ 
                                                   &                           &  7/2 & 384.7872      & 0.0003  &   60 \\ 
                                                   &                           &  5/2 & 667.6829      & 0.0003  &   64 \\ 
                                                   &                           &  3/2 & 862.6118      & 0.0004  &   52 \\ 
                                                   &                           &  1/2 & 977.0498      & 0.0004  &   30 \\ 
                                 $\mathrm{3d^7 }$  &         $\mathrm{ a^4F }$ &  9/2 & 1872.5998     & 0.0006  &   63 \\ 
                                                   &                           &  7/2 & 2430.1369     & 0.0006  &   77 \\ 
                                                   &                           &  5/2 & 2837.9807     & 0.0007  &   76 \\
\hline                  
\end{longtable}

\begin{landscape}
\setlength\LTleft{-0.5in}
\setlength\LTright{-1in plus 1 fill}
\scriptsize                           
\setcounter{table}{3}
\begin{longtable}{rddddddddr@{}r@{ - }r@{}rl}   
\caption{Observed lines of Fe II} \\ 
\\ \hline  Intensity$^a$ 
      & \multicolumn{1}{c}{ Vacuum }          
      & \multicolumn{1}{c}{Unc.$^b$   }       
      & \multicolumn{1}{c}{Wavenumber}    
      & \multicolumn{1}{c}{Unc.$^b$       }   
      & \multicolumn{1}{c}{Ritz Vacuum }      
      & \multicolumn{1}{c}{Unc.$^b$         } 
      & \multicolumn{1}{c}{Lower level      } 
      & \multicolumn{1}{c}{Upper level      } 
      & \multicolumn{4}{c}{Classification } 
      & Notes$^c$ \\           
      &   \multicolumn{1}{c}{wavelength}  & & & & \multicolumn{1}{c}{wavelength} \\ 
      & \multicolumn{1}{c}{(\AA\ )   }                      
      & \multicolumn{1}{c}{(\AA\ )    }                     
      & \multicolumn{1}{c}{($ \mathrm{cm^{-1}}$) }           
      & \multicolumn{1}{c}{($  \mathrm{cm^{-1}}$) } 
      & \multicolumn{1}{c}{($ \mathrm{\AA\ }$) }  
      & \multicolumn{1}{c}{($ \mathrm{\AA\ }$) }  
      & \multicolumn{1}{c}{($  \mathrm{cm^{-1}}$) } 
      & \multicolumn{1}{c}{($  \mathrm{cm^{-1}}$) } 
      & \multicolumn{3}{c}{                     }     \\  
\\ \hline                              
 \endfirsthead 
 \multicolumn{6}{l} { \tablename\ \thetable{} Observed lines of Fe~II  -- continued from previous page} \\ 
\\ \hline  Intensity$^a$ 
      & \multicolumn{1}{c}{ Vacuum }               
      & \multicolumn{1}{c}{Unc.$^b$   }              
      & \multicolumn{1}{c}{Wavenumber}           
      & \multicolumn{1}{c}{Unc.$^b$       }          
      & \multicolumn{1}{c}{Ritz Vacuum }      
      & \multicolumn{1}{c}{Unc.$^b$         }        
      & \multicolumn{1}{c}{Lower level      } 
      & \multicolumn{1}{c}{Upper level      } 
      & \multicolumn{4}{c}{Classification }      
      & Notes$^c$ \\ 
      &  \multicolumn{1}{c}{wavelength}  & & & & \multicolumn{1}{c}{wavelength} \\ 
      & \multicolumn{1}{c}{(\AA\ )   }                      
      & \multicolumn{1}{c}{(\AA\ )    }                     
      & \multicolumn{1}{c}{($ \mathrm{cm^{-1}}$) }           
      & \multicolumn{1}{c}{($  \mathrm{cm^{-1}}$) } 
      & \multicolumn{1}{c}{($ \mathrm{\AA\ }$) }  
      & \multicolumn{1}{c}{($ \mathrm{\AA\ }$) }  
      & \multicolumn{1}{c}{($  \mathrm{cm^{-1}}$) } 
      & \multicolumn{1}{c}{($  \mathrm{cm^{-1}}$) } 
      & \multicolumn{3}{c}{                     }               \\  
\\ \hline                              
 \endhead 
\hline \multicolumn{12}{r}{{Continued on next page}} \\ 
\endfoot 
\hline 
\endlastfoot 
                   8  &  49777.40     & 0.03     & 2008.9438    & 0.0011   & 49777.42     & 0.04     & 104174.285  & 106183.228  &  $\mathrm{3d^6(^3H)4d }$  & $\mathrm{ ^4I  }$$_{11/2 }$ &  $\mathrm{3d^5(^2I)\,4s4p(^3P) }$  & $\mathrm{^2K^{\circ}}$$_{13/2 }$ &      \\ 
                   4  &  49745.78     & 0.06     & 2010.221     & 0.002    & 49745.728    & 0.011    & 105449.557  & 107459.780  &  $\mathrm{3d^6(^5D)5d }$  & $\mathrm{^4G}$$_{7/2 }$ &  $\mathrm{3d^6(^5D)6p }$  & $\mathrm{^6F^{\circ}}$$_{7/2 }$ &      \\ 
                   4  &  49430.03     & 0.11     & 2023.062     & 0.004    & 49429.86     & 0.03     & 104569.226  & 106592.294  &  $\mathrm{3d^6(^3P2)4d }$  & $\mathrm{ ^4F  }$$_{5/2 }$ &  $\mathrm{3d^5(^4F)\,4s4p(^3P) }$  & $\mathrm{^6F^{\circ} }$$_{5/2 }$ &   *   I\\ 
                   3  &  49278.41     & 0.07     & 2029.286     & 0.003    & 49278.444    & 0.012    & 104863.432  & 106892.717  &  $\mathrm{3d^6(^5D)5d }$  & $\mathrm{^4G}$$_{11/2 }$ &  $\mathrm{3d^6(^5D)6p }$  & $\mathrm{^6F^{\circ}}$$_{11/2 }$ &      \\ 
                  26  &  49228.588    & 0.009    & 2031.3400    & 0.0004   & 49228.617    & 0.011    & 104593.292  & 106624.631  &  $\mathrm{3d^6(^5D)5d }$  & $\mathrm{^6G}$$_{11/2 }$ &  $\mathrm{3d^5(^4F)\,4s4p(^3P) }$  & $\mathrm{^6F^{\circ} }$$_{9/2 }$ &      \\ 
                   8  &  49113.52     & 0.03     & 2036.0991    & 0.0014   & 49113.545    & 0.010    & 105127.773  & 107163.871  &  $\mathrm{3d^6(^5D)5d }$  & $\mathrm{^4D}$$_{5/2 }$ &  $\mathrm{3d^6(^5D)6p }$  & $\mathrm{^4P^{\circ}}$$_{5/2 }$ &      \\ 
                   3  &  49102.3      & 0.2      & 2036.565     & 0.010    & 49102.242    & 0.013    & 112994.097  & 115030.664  &  $\mathrm{3d^6(^5D_4)5f }$  & $\mathrm{[6]^{\circ}}$$_{11/2 }$ &  $\mathrm{3d^6(^3D)4d }$  & $\mathrm{  ^4F        }$$_{9/2 }$ &      \\
\hline 
  \end{longtable} 
\setcounter{table}{3}
\begin{longtable}{rddddddddr@{}r@{ - }r@{}rl}   
\\ \hline  Intensity$^a$ 
      & \multicolumn{1}{c}{ Air }          
      & \multicolumn{1}{c}{Unc.$^b$   }       
      & \multicolumn{1}{c}{Wavenumber}    
      & \multicolumn{1}{c}{Unc.$^b$       }   
      & \multicolumn{1}{c}{Ritz Air }      
      & \multicolumn{1}{c}{Unc.$^b$         } 
      & \multicolumn{1}{c}{Lower level      } 
      & \multicolumn{1}{c}{Upper level      } 
      & \multicolumn{4}{c}{Classification } 
      & Notes$^c$ \\           
      &   \multicolumn{1}{c}{wavelength}  & & & & \multicolumn{1}{c}{wavelength} \\ 
      & \multicolumn{1}{c}{(\AA\ )   }                      
      & \multicolumn{1}{c}{(\AA\ )    }                     
      & \multicolumn{1}{c}{($ \mathrm{cm^{-1}}$) }           
      & \multicolumn{1}{c}{($  \mathrm{cm^{-1}}$) } 
      & \multicolumn{1}{c}{($ \mathrm{\AA\ }$) }  
      & \multicolumn{1}{c}{($ \mathrm{\AA\ }$) }  
      & \multicolumn{1}{c}{($  \mathrm{cm^{-1}}$) } 
      & \multicolumn{1}{c}{($  \mathrm{cm^{-1}}$) } 
      & \multicolumn{3}{c}{                     }     \\  
\\ \hline                              
 \endfirsthead 
 \multicolumn{6}{l} { \tablename\ \thetable{} Observed lines of Fe~II  -- continued from previous page} \\ 
\\ \hline  Intensity$^a$ 
      & \multicolumn{1}{c}{ Air }               
      & \multicolumn{1}{c}{Unc.$^b$   }              
      & \multicolumn{1}{c}{Wavenumber}           
      & \multicolumn{1}{c}{Unc.$^b$       }          
      & \multicolumn{1}{c}{Ritz Air }      
      & \multicolumn{1}{c}{Unc.$^b$         }        
      & \multicolumn{1}{c}{Lower level      } 
      & \multicolumn{1}{c}{Upper level      } 
      & \multicolumn{4}{c}{Classification }      
      & Notes$^c$ \\ 
      &  \multicolumn{1}{c}{wavelength}  & & & & \multicolumn{1}{c}{wavelength} \\ 
      & \multicolumn{1}{c}{(\AA\ )   }                      
      & \multicolumn{1}{c}{(\AA\ )    }                     
      & \multicolumn{1}{c}{($ \mathrm{cm^{-1}}$) }           
      & \multicolumn{1}{c}{($  \mathrm{cm^{-1}}$) } 
      & \multicolumn{1}{c}{($ \mathrm{\AA\ }$) }  
      & \multicolumn{1}{c}{($ \mathrm{\AA\ }$) }  
      & \multicolumn{1}{c}{($  \mathrm{cm^{-1}}$) } 
      & \multicolumn{1}{c}{($  \mathrm{cm^{-1}}$) } 
      & \multicolumn{3}{c}{                     }               \\  
\\ \hline                              
 \endhead 
\hline \multicolumn{12}{r}{{Continued on next page}} \\ 
\endfoot 
\hline 
\endlastfoot 
                  37  &  19981.4909   & 0.0014   & 5003.2657    & 0.0003   & 19981.465    & 0.004    &  94685.083  &  99688.355  &  $\mathrm{3d^5(^4D)\,4s4p(^3P) }$  & $\mathrm{^6P^{\circ} }$$_{5/2 }$ &  $\mathrm{3d^6(^3F2)5s }$  & $\mathrm{f^4F  }$$_{7/2 }$ &   *   I\\ 
                  61  &  19977.6306   & 0.0013   & 5004.2325    & 0.0003   & 19977.6330   & 0.0019   &  84326.967  &  89331.199  &  $\mathrm{3d^6(^5D)4d }$  & $\mathrm{^6P  }$$_{5/2 }$ &  $\mathrm{3d^6(^5D)5p }$  & $\mathrm{^6D^{\circ}}$$_{3/2 }$ &      \\ 
                 217  &  19964.9548   & 0.0011   & 5007.4097    & 0.0003   & 19964.9568   & 0.0014   & 102030.965  & 107038.375  &  $\mathrm{3d^6(^5D)6s }$  & $\mathrm{^6D }$$_{7/2 }$ &  $\mathrm{3d^5(^2D)\,4s4p(^3P) }$  & $\mathrm{^4D^{\circ}}$$_{7/2 }$ &      \\ 
                   3  &  19959.50     & 0.05     & 5008.779     & 0.013    & 19959.477    & 0.005    & 108365.344  & 113374.128  &  $\mathrm{3d^6(^3G)4d }$  & $\mathrm{ ^4D}$$_{7/2 }$ &  $\mathrm{3d^6(^5D_3)5f }$  & $\mathrm{[4]^{\circ}}$$_{7/2 }$ &      \\ 
                 114  &  19955.3680   & 0.0010   & 5009.8153    & 0.0002   & 19955.3685   & 0.0019   &  83713.583  &  88723.398  &  $\mathrm{3d^6(^5D)4d }$  & $\mathrm{^6D  }$$_{7/2 }$ &  $\mathrm{3d^6(^5D)5p }$  & $\mathrm{^6D^{\circ}}$$_{9/2 }$ &      \\ 
                  13  &  19950.726    & 0.005    & 5010.9810    & 0.0012   & 19950.736    & 0.003    & 103921.637  & 108932.616  &  $\mathrm{3d^6(^3H)4d }$  & $\mathrm{ ^4G  }$$_{7/2 }$ &  $\mathrm{3d^6(^3P2)5p }$  & $\mathrm{ ^2D^{\circ} }$$_{5/2 }$ &      \\ 
                   2  &  19948.878    & 0.012    & 5011.445     & 0.003    & 19948.892    & 0.002    & 105155.093  & 110166.534  &  $\mathrm{3d^6(^3F2)4d }$  & $\mathrm{^4G  }$$_{9/2 }$ &  $\mathrm{3d^6(^3F)5p }$  & $\mathrm{ ^4F^{\circ}}$$_{9/2 }$ &      \\
\hline 
  \end{longtable} 
\setcounter{table}{3}
\begin{longtable}{rddddddddr@{}r@{ - }r@{}rl}   
\\ \hline  Intensity$^a$ 
      & \multicolumn{1}{c}{ Vacuum }          
      & \multicolumn{1}{c}{Unc.$^b$   }       
      & \multicolumn{1}{c}{Wavenumber}    
      & \multicolumn{1}{c}{Unc.$^b$       }   
      & \multicolumn{1}{c}{Ritz Vacuum }      
      & \multicolumn{1}{c}{Unc.$^b$         } 
      & \multicolumn{1}{c}{Lower level      } 
      & \multicolumn{1}{c}{Upper level      } 
      & \multicolumn{4}{c}{Classification } 
      & Notes$^c$ \\           
      &   \multicolumn{1}{c}{wavelength}  & & & & \multicolumn{1}{c}{wavelength} \\ 
      & \multicolumn{1}{c}{(\AA\ )   }                      
      & \multicolumn{1}{c}{(\AA\ )    }                     
      & \multicolumn{1}{c}{($ \mathrm{cm^{-1}}$) }           
      & \multicolumn{1}{c}{($  \mathrm{cm^{-1}}$) } 
      & \multicolumn{1}{c}{($ \mathrm{\AA\ }$) }  
      & \multicolumn{1}{c}{($ \mathrm{\AA\ }$) }  
      & \multicolumn{1}{c}{($  \mathrm{cm^{-1}}$) } 
      & \multicolumn{1}{c}{($  \mathrm{cm^{-1}}$) } 
      & \multicolumn{3}{c}{                     }     \\  
\\ \hline                              
 \endfirsthead 
 \multicolumn{6}{l} { \tablename\ \thetable{} Observed lines of Fe~II  -- continued from previous page} \\ 
\\ \hline  Intensity$^a$ 
      & \multicolumn{1}{c}{ Vacuum }               
      & \multicolumn{1}{c}{Unc.$^b$   }              
      & \multicolumn{1}{c}{Wavenumber}           
      & \multicolumn{1}{c}{Unc.$^b$       }          
      & \multicolumn{1}{c}{Ritz Vacuum }      
      & \multicolumn{1}{c}{Unc.$^b$         }        
      & \multicolumn{1}{c}{Lower level      } 
      & \multicolumn{1}{c}{Upper level      } 
      & \multicolumn{4}{c}{Classification }      
      & Notes$^c$ \\ 
      &  \multicolumn{1}{c}{wavelength}  & & & & \multicolumn{1}{c}{wavelength} \\ 
      & \multicolumn{1}{c}{(\AA\ )   }                      
      & \multicolumn{1}{c}{(\AA\ )    }                     
      & \multicolumn{1}{c}{($ \mathrm{cm^{-1}}$) }           
      & \multicolumn{1}{c}{($  \mathrm{cm^{-1}}$) } 
      & \multicolumn{1}{c}{($ \mathrm{\AA\ }$) }  
      & \multicolumn{1}{c}{($ \mathrm{\AA\ }$) }  
      & \multicolumn{1}{c}{($  \mathrm{cm^{-1}}$) } 
      & \multicolumn{1}{c}{($  \mathrm{cm^{-1}}$) } 
      & \multicolumn{3}{c}{                     }               \\  
\\ \hline                              
 \endhead 
\hline \multicolumn{12}{r}{{Continued on next page}} \\ 
\endfoot 
\hline 
\endlastfoot 
                   8  &  1999.4599    & 0.0006   & 50013.505    & 0.016    & 1999.46003   & 0.00009  &  22637.1950 &  72650.698  &  $\mathrm{3d^6(^3F2)4s }$  & $\mathrm{b^4F  }$$_{9/2 }$ &  $\mathrm{3d^6(^3D)4p }$  & $\mathrm{  w^4F^{\circ}}$$_{9/2 }$ & F    \\ 
                  17  &  1999.4114    & 0.0003   & 50014.718    & 0.008    & 1999.41143   & 0.00009  &  22637.1950 &  72651.914  &  $\mathrm{3d^6(^3F2)4s }$  & $\mathrm{b^4F  }$$_{9/2 }$ &  $\mathrm{3d^6(^3D)4p }$  & $\mathrm{  w^4D^{\circ}}$$_{7/2 }$ & F * III\\ 
                  0?  &  1999.342     & 0.005    & 50016.46     & 0.13     & 1999.35341   & 0.00009  &  63465.134  & 113481.304  &  $\mathrm{3d^6(^3F2)4p }$  & $\mathrm{x^4D^{\circ} }$$_{3/2 }$ &  $\mathrm{3d^6(^5D)7s }$  & $\mathrm{^4D }$$_{3/2 }$ & N    \\ 
                   2. &  1999.206     & 0.005    & 50019.87     & 0.13     & 1999.20603   & 0.00020  &  22409.8178 &  72429.675  &  $\mathrm{3d^6(^3P2)4s }$  & $\mathrm{b^4P  }$$_{1/2 }$ &  $\mathrm{3d^6(^3D)4p }$  & $\mathrm{  w^4D^{\circ}}$$_{1/2 }$ & A    \\ 
                 0d?  &  1999.052     & 0.005    & 50023.71     & 0.13     & 1999.05524   & 0.00011  &  23031.2829 &  73054.913  &  $\mathrm{3d^6(^3F2)4s }$  & $\mathrm{b^4F  }$$_{3/2 }$ &  $\mathrm{3d^6(^1G2)4p }$  & $\mathrm{x^2F^{\circ} }$$_{5/2 }$ & N  II\\ 
                 0d?  &  1999.052     & 0.005    & 50023.71     & 0.13     & 1999.04759   & 0.00009  &  60445.279  & 110469.101  &  $\mathrm{3d^5\,4s^2 }$  & $\mathrm{c^4D }$$_{5/2 }$ &  $\mathrm{3d^6(^3F)5p }$  & $\mathrm{ ^4D^{\circ}}$$_{7/2 }$ & N  II\\ 
                   4  &  1998.9360    & 0.0007   & 50026.614    & 0.017    & 1998.93508   & 0.00009  &  57578.490  & 107605.127  &  $\mathrm{3d^5\,4s^2 }$  & $\mathrm{d^4P }$$_{1/2 }$ &  $\mathrm{3d^6(^5D)6p }$  & $\mathrm{^6F^{\circ}}$$_{3/2 }$ & F    \\
\hline 
\end{longtable}
\noindent
$^a$ Intensities of the lines with different codes in `Notes' column
are on different scale and should not be relied on for accurate intensity
measurements. Decimal numbers are from grating 
spectra. Integer numbers are from FT spectra, which are in general the 
stronger lines. \\ 
$^b$ One standard uncertainty of preceding column.\\  
$^c$ *: Line given low weight in level optimization; 
     F: Line measured in FT spectra; 
     G: Line measured in grating spectra; 
     N: Line is from unpublished iron-neon linelists (see section 3.2). 
     A: Line is from unpublished iron-argon linelists (see section 3.2). 
     L: Line is from unpublished linelists below 1600~AA (see section 3.2). 
     S: Line measured in grating spectra and is saturated on plate.
     I, II, III : Line blended with Fe~I, Fe~II, or Fe~III. 
     Ne, Ar, Si, Ti, Cr, Co, Ni  : Line blended with species indicated. 
     gh  : Line possibly blended with ghost. 
\end{landscape}

\begin{harvard}
\expandafter\ifx\csname natexlab\endcsname\relax\def\natexlab#1{#1}\fi

\bibitem[{Adam {et~al.}(1987)Adam, Baschek, Johansson, Nilsson, \&
  Brage}]{Adam_87}
Adam, J., Baschek, B., Johansson, S., Nilsson, A.~E., \& Brage, T. 1987,
  Astrophys. J., 312, 337

\bibitem[{Aldenius(2009)}]{Aldenius_09}
Aldenius, M. 2009, Phys. Scr., T134, 014008

\bibitem[{Aldenius {et~al.}(2006)Aldenius, Johansson, \& Murphy}]{Aldenius_06}
Aldenius, M., Johansson, S.~G., \& Murphy, M.~T. 2006, Mon. Not. R. Astron.
  Soc., 370, 444

\bibitem[{Batteiger {et~al.}(2009)Batteiger, Kn\"unz, Herrmann, Saathoff,
  Sch\"ussler, Bernhardt, Wilken, Holzwarth, H\"ansch, \& Udem}]{Batteiger_09}
Batteiger, V., Kn\"unz, S., Herrmann, M., {et~al.} 2009, Phys. Rev. A, 80,
  022503

\bibitem[{Bi\'emont {et~al.}(1997)Bi\'emont, Johansson, \&
  Palmeri}]{Biemont_97}
Bi\'emont, E., Johansson, S., \& Palmeri, P. 1997, Phys. Scr., 55, 559

\bibitem[{{Brandt} {et~al.}(1999){Brandt}, {Heap}, {Beaver}, {Boggess},
  {Carpenter}, {Ebbets}, {Hutchings}, {Jura}, {Leckrone}, {Linsky}, {Maran},
  {Savage}, {Smith}, {Trafton}, {Walter}, {Weymann}, {Proffitt}, {Wahlgren},
  {Johansson}, {Nilsson}, {Brage}, {Snow}, \& {Ake}}]{chilupi}
{Brandt}, J.~C., {Heap}, S.~R., {Beaver}, E.~A., {et~al.} 1999, Astron. J.,
  117, 1505

\bibitem[{Brault \& Abrams(1989)}]{Brault_89}
Brault, J.~W., \& Abrams, M.~C. 1989, in {F}ourier Transform Spectroscopy:New
  Methods and Applications, Vol.~6, 110--112

\bibitem[{Castelli \& Kurucz(2010)}]{Castelli_10}
Castelli, F., \& Kurucz, R.~L. 2010, Astron. Astrophys., 520, p. A57

\bibitem[{{Castelli} {et~al.}(2009){Castelli}, {Kurucz}, \&
  {Hubrig}}]{Castelli_08}
{Castelli}, F., {Kurucz}, R.~L., \& {Hubrig}, S. 2009, Astron. Astrophys., 508,
  401

\bibitem[{Davis {et~al.}(2001)Davis, Abrams, \& Brault}]{Davis_01}
Davis, S.~P., Abrams, M.~C., \& Brault, J.~W. 2001, {F}ourier transform
  spectrometry (Academic Press)

\bibitem[{{Dupree} {et~al.}(2005){Dupree}, {Lobel}, {Young}, {Ake}, {Linsky},
  \& {Redfield}}]{Dupree_05}
{Dupree}, A.~K., {Lobel}, A., {Young}, P.~R., {et~al.} 2005, Astrophys. J.,
  622, 629

\bibitem[{{Hamann} \& {Ferland}(1999)}]{Hamann_1999}
{Hamann}, F., \& {Ferland}, G. 1999, Ann. Review Astron. Astrophys., 37, 487

\bibitem[{Hannemann {et~al.}(2006)Hannemann, Salumbides, Witte, Zinkstok, van
  Duijn, Eikema, \& Ubachs}]{Hannemann_06}
Hannemann, S., Salumbides, E.~J., Witte, S., {et~al.} 2006, Phys. Rev. A, 74,
  012505

\bibitem[{{Hill} \& {Suarez}(2012)}]{NSO}
{Hill}, F., \& {Suarez}, I. 2012, National Solar Observatory Digital Library,
  {\em{http://diglib.nso.edu}}

\bibitem[{Johansson(1978)}]{Johansson_78}
Johansson, S. 1978, Phys. Scr., 18, 217

\bibitem[{Johansson(2009)}]{Johansson_09}
---. 2009, Phys. Scr., T134, 014013

\bibitem[{Johansson \& Baschek(1988)}]{Johansson_88}
Johansson, S., \& Baschek, B. 1988, Nucl. Instrum. Methods Phys. Res. B, 31,
  222

\bibitem[{Johansson {et~al.}(1995)Johansson, Brage, Leckrone, Nave, \&
  Wahlgren}]{Johansson_95}
Johansson, S., Brage, T., Leckrone, D.~S., Nave, G., \& Wahlgren, G.~M. 1995,
  Astrophys. J., 446, 361

\bibitem[{Johansson \& Litz\'en(1974)}]{Johansson_74}
Johansson, S., \& Litz\'en, U. 1974, Phys. Scr., 10, 121

\bibitem[{Kramida(2011)}]{Kramida_11}
Kramida, A.~E. 2011, Comput. Phys. Commun., 182, 419

\bibitem[{Kramida \& Nave(2006)}]{Kramida_06}
Kramida, A.~E., \& Nave, G. 2006, Eur. Phys. J. D, 39, 331

\bibitem[{Kurucz(2010)}]{Kurucz_10}
Kurucz, R. 2010, Atomic line data for {F}e {II}: file gf2601.pos created on
  23rd July, 2010, {\em{http://kurucz.harvard.edu/atoms/2601/gf2601.pos }}

\bibitem[{Learner \& Thorne(1988)}]{Learner_88}
Learner, R. C.~M., \& Thorne, A.~P. 1988, J. Opt. Soc. Am. B, 5, 2045

\bibitem[{Nave {et~al.}(1994)Nave, Johansson, Learner, Thorne, \&
  Brault}]{Nave_94}
Nave, G., Johansson, S., Learner, R. C.~M., Thorne, A.~P., \& Brault, J.~W.
  1994, Astrophys. J., Suppl. Ser., 94, 221

\bibitem[{Nave {et~al.}(1997{\natexlab{a}})Nave, Johansson, \&
  Thorne}]{Nave_97}
Nave, G., Johansson, S., \& Thorne, A.~P. 1997{\natexlab{a}}, J. Opt. Soc. Am.
  B, 14, 1035

\bibitem[{Nave {et~al.}(1992)Nave, Learner, Murray, Thorne, \&
  Brault}]{Nave_92}
Nave, G., Learner, R. C.~M., Murray, J.~E., Thorne, A.~P., \& Brault, J.~W.
  1992, J. Phys. II (France), 2, 913

\bibitem[{Nave {et~al.}(1991)Nave, Learner, Thorne, \& Harris}]{Nave_91}
Nave, G., Learner, R. C.~M., Thorne, A.~P., \& Harris, C.~J. 1991, J. Opt. Soc.
  Am. B, 8, 2028

\bibitem[{Nave \& Sansonetti(2004)}]{Nave_04}
Nave, G., \& Sansonetti, C.~J. 2004, J. Opt. Soc. Am. B, 21, 442

\bibitem[{Nave \& Sansonetti(2011)}]{Nave_11}
---. 2011, J. Opt. Soc. Am. B, 28, 737

\bibitem[{Nave {et~al.}(1997{\natexlab{b}})Nave, Sansonetti, \&
  Griesmann}]{Nave_97b}
Nave, G., Sansonetti, C.~J., \& Griesmann, U. 1997{\natexlab{b}}, in {F}ourier
  Transform Spectroscopy:New Methods and Applications, Vol.~3, 38--40

\bibitem[{Peck \& Reeder(1972)}]{Peck_72}
Peck, E.~R., \& Reeder, K. 1972, J. Opt. Soc. Am., 62, 958

\bibitem[{{Ralchenko} {et~al.}(2012){Ralchenko}, {Kramida}, {Reader}, \& {NIST
  ASD Team}}]{ASD}
{Ralchenko}, Y., {Kramida}, A., {Reader}, J., \& {NIST ASD Team}. 2012, NIST
  Atomic Spectra Database (version 4.1), {\em{http://physics.nist.gov/asd}}

\bibitem[{Rosberg \& Johansson(1992)}]{Rosberg_92}
Rosberg, M., \& Johansson, S. 1992, Phys. Scr., 45, 590

\bibitem[{{S. Johansson} \& {V. S. Letokhov}(2004)}]{Johansson_lasers}
{S. Johansson}, \& {V. S. Letokhov}. 2004, A\&A, 428, 497

\bibitem[{Saloman \& Sansonetti(2004)}]{Saloman_04}
Saloman, E.~B., \& Sansonetti, C.~J. 2004, J. Phys. Chem. Ref. Data, 33, 1113

\bibitem[{Salumbides {et~al.}(2006)Salumbides, Hannemann, Eikema, \&
  Ubachs}]{Salumbides_06}
Salumbides, E.~J., Hannemann, S., Eikema, K. S.~E., \& Ubachs, W. 2006, Mon.
  Not. R. Astron. Soc., 373, L41

\bibitem[{Sansonetti {et~al.}(2004)Sansonetti, Blackwell, \&
  Saloman}]{Sansonetti_04}
Sansonetti, C.~J., Blackwell, M.~M., \& Saloman, E.~B. 2004, J. Res. Natl.
  Inst. Stand. Technol., 109, 371

\bibitem[{Schoenfeld {et~al.}(1995)Schoenfeld, Chang, Geller, Johansson, Nave,
  Sauval, \& Grevesse}]{Schoenfeld_95}
Schoenfeld, W.~G., Chang, E.~S., Geller, M., {et~al.} 1995, Astron. Astrophys.,
  301, 593

\bibitem[{Whaling {et~al.}(1995)Whaling, Anderson, Carle, Brault, \&
  Zarem}]{Whaling_95}
Whaling, W., Anderson, W. H.~C., Carle, M.~T., Brault, J.~W., \& Zarem, H.~A.
  1995, J. Quant. Spectrosc. Radiat. Transfer, 53, 1

\bibitem[{Whaling {et~al.}(2002)Whaling, Anderson, Carle, Brault, \&
  Zarem}]{Whaling_07}
---. 2002, J. Res. Natl. Inst. Stand. Technol., 107, 149

\end{harvard}
\end{document}